\documentclass[12pt]{iopart}

\usepackage{braket,graphicx}
\usepackage{iopams}  
\begin{document}

\title[Density operator of a system pumped with polaritons: A Jaynes-Cummings
like approach]{Density operator of a system pumped with polaritons: A
Jaynes-Cummings
like approach}

\author{Nicol\'as Quesada$^\dagger$, Herbert Vinck-Posada$^*$ and Boris A.
Rodr\'iguez$^\dagger$}

\address{$^\dagger$ Instituto de F\'isica, Universidad de Antioquia,
    Medell\'in, AA 1226 Medell\'in, Colombia \\
$^*$ Departamento de F\'isica, Universidad Nacional de
Colombia, Ciudad Universitaria, Bogot\'a, Colombia
}
\ead{nquesada@pegasus.udea.edu.co}
\begin{abstract}
We investigate the effects of considering two different incoherent pumpings over
a microcavity-quantum dot system modelled using the Jaynes-Cummings
Hamiltonian. When the system is  incoherently pumped with polaritons it is able
to sustain a large number of photons inside the cavity with Poisson-like
statistics in the stationary limit, and also leads to a separable
exciton-photon state. We also investigate the effects of both types of pumpings
(Excitonic and Polaritonic) in the emission spectrum of the cavity. We show that the 
polaritonic pumping as considered here is unable to modify the dynamical regimes of the 
system as the excitonics pumping does. Finally, we obtain a closed form expression 
for the negativity of the density
matrices that the quantum master equation considered here generates.

\end{abstract}

%Uncomment for PACS numbers title message
% Keywords required only for MST, PB, PMB, PM, JOA, JOB? 
%\vspace{2pc}
%\noindent{\it Keywords}: Article preparation, IOP journals
% Uncomment for Submitted to journal title message
\submitto{\JPCM}
% Comment out if separate title page not required
\maketitle

\section{Introduction}
In the last few years the study of solid-state systems of nanometric dimension
in which light interacts with matter has become a widely studied subject. 
The study of excitons interacting with a confined mode of light has made
possible the observation of two different coupling regimes  \cite{Laussy-book,
Yamamoto-book} (strong and weak coupling). It also made possible the observation
of collective phenomena of quasiparticles in semiconductor microcavitites
\cite{coh1,coh2,coh3}. Motivated by very interesting experimental results
regarding condensation of polaritons \cite{Deng06,LeSiDang} it has been proposed
that some of the coherence properties of the above mentioned system can be
understood in terms of an
effective pump of polaritons \cite{Vera09, Vera09b}. In these  references, a
finite system of electrons and holes confined in a parabolic Quantum Dot (QD)
is used to model the matter component of the system. They first obtain the
dressed states of the finite system model (using a numerical diagonalization
procedure), in which the Coulomb interaction between charge carriers and the
dipole interaction between light and matter is explicitly included. Then they
use a zero temperature quantum master equation in which they include besides the
relaxation processes due to coherent emission, an incoherent pump of polaritons.
The theoretical results they obtain, with this model, reproduce the polariton
laser threshold reported in \cite{BlochJ}.\\
In a recent theoretical and experimental study \cite{Finley}, using a simple
Jaynes-Cummings like model including dissipative processes, the authors obtain a
surprisingly good agreement between the calculated and measured polariton
spectra. They show that the effective dissipative parameters of the system
depend on the nominal excitation power density. They also discuss the
difficulties involved in determining the strong coupling (SC) regime mainly
because the broadening of the spectral lines hinders the well known
anti-crossing characteristic feature. The determination of a clear signature of
the strong coupling regime motivated several recent theoretical works
\cite{teje1,teje2,teje3}.\\
The above paragraph highlights the importance of studying the role of the
incoherent pumping in the determination of the dynamical regimes in the system.
In particular, the effect of an incoherent pumping of polaritons or excitons must
be clarified. In this work we seek to discuss the dynamical effects of the
interplay between polaritonic-excitonic pumping, by using the simplest model of
quantized light matter interaction, the Jaynes-Cummings model
\cite{jaynes,Gerry,Tejedor04}. We also seek to understand how the entanglement
between the excitons and photons is affected by the incoherent pumping since we
have two strongly coupled interacting systems. Some previous works in this
direction are \cite{Vera09JPCM,Tejedor07,ElenaA,ElenaArxiv}.\\ 
The paper has been written as
follows: In section 2 a description of the master equation of the system is
given. The matrix elements of the density operator related to the polariton
pumping Lindblad superoperator are obtained. We also show the operators and
dynamical equations necessary to obtain the emission spectrum using the quantum
regression theorem. Finally, in this section we derive a closed form expression
for the Peres positive partial transpose criterion in order to quantify the
exciton-photon entanglement. In section 3 we compare the effects of both types
of pumping in the statistical properties of the steady state density operator.
Then we show, analytically and numerically, how the polariton pumping affects
the emission spectra of the system. Finally, in this section it is shown the
effects of both types of pumping in the entanglement of the system.

\section{Theoretical Background}
We are interested in studying an exciton interacting with the lowest energy
(frequency) mode of a semiconductor microcavity. The quantum states 
resultant of the electrostatic interaction between holes in the valence band
and electrons in the conduction band in a solid state system are termed
excitons. This quasiparticles
might exhibit a discrete or continuum spectrum depending on their confinement.
In this work we shall consider only the lowest energy levels of the system, the
ground state, $\ket{G}$  (no excitation, \emph{i.e.} electron in the valence
band) and excited $\ket{X}$ state. In this effective formulation all the
complexities related to the many body problem of considering electrons and holes
in quantum dot or well are effectively included in the energy separation
between the ground and exciton states. The light will be treated as a single
electromagnetic quantized mode. This assumption is subjected to existence of well
separated energy modes in the cavity we are considering. This assumptions are
the ones usually considered in most theoretical work
\cite{Finley,teje1,Tejedor04,Vera09JPCM}. 

\subsection{Hamiltonian and Dressed States}
The Hamiltonian we shall use to study this system is the well known
Jaynes-Cummings Hamiltonian \cite{jaynes}:
\begin{equation}
 H=(\omega_X-\Delta)a^\dagger a+\omega_X
\sigma^\dagger \sigma+g(\sigma  a^\dagger+\sigma^{\dagger} a),
\label{HJC}
\end{equation}
where $\ket{X}$,$\ket{G}$ are the exciton and ground states of the matter,
$\sigma=\ket{G}\bra{X}$, $\sigma^{\dagger}=\ket{X}\bra{G}$ 
, $a$,$a^\dagger$ are the annihilation and creation operators of the field, $\omega_X$ is the 
energy required to create an exciton, $\Delta=\omega_X-\omega_0$ is the detuning between the exciton and photon 
frequencies, $g$ is the light-matter coupling constant or Rabi constant
and we have taken $\hbar \equiv 1$. \\
This Hamiltonian is obtained after considering that the exciton is coupled only
to one mode of the microcavity, that the interaction between them is of dipole
type (\emph{i.e.} the spatial variation of the electromagnetic field is small in the
spatial dimensions of the exciton) and after neglecting counter rotating terms
(the so called Rotating Wave Approximation) \cite{Gerry}. 

The Hamiltonian $H$ can be diagonalized when written in the basis $\{
\ket{G},\ket{X} \} \otimes \{ \ket{n} \} _{n=0}^{\infty}$ (The Bared Basis).
It takes the block-diagonal form:
\begin{equation}
 \left(
\begin{array}{ll}
 (n-1) (\omega_X-\Delta) +\omega _X & g \sqrt{n} \\
 g \sqrt{n} & n (\omega_X-\Delta)
\end{array}
\right),
\end{equation}
when written in the $n$-th excitation manifold basis $\{\ket{Xn-1},\ket{Gn} \}$. The
eigenvalues
and eigenvectors can be easily obtained and are given by \cite{Gerry}:
\begin{eqnarray}
\omega_{n\pm}=\frac{1}{2} \left(-2 n \Delta +\Delta \pm \sqrt{4 n
g^2+\Delta
   ^2}\right)+n \omega _X\\
\left(
\begin{array}{l}
 \ket{n,+} \\
 \ket{n,-}
\end{array}
\right) =
A \left(
\begin{array}{l}
 \ket{X n-1} \\
 \ket{G n}
\end{array}
\right).
\end{eqnarray}
Where $A$ is clock wise rotation matrix by the angle
$\phi_n/2=\tan^{-1}\left( \frac{g\sqrt{n}}{\Delta}\right)/2$.
The states $\ket{n, \pm}$ are the dressed states of the Hamiltonian (\ref{HJC}).
If the states $\ket{G}, \ket{X}$ are the excitonic states of the QD then
the states $\ket{n, \pm}$ can be considered the polaritonic states of the
system.

The emission spectrum of the system in this idealistic model is given by the
transitions that can occur between two given dressed states and the values of
the transition energies are precisely the differences of their respective
frequencies times $\hbar$.\\

\subsection{Master Equation \label{popelem}}
\begin{figure}[htb]
\centering
 \begin{minipage}{0.48\textwidth}
  \centering a)
\includegraphics[width=1.0\textwidth]{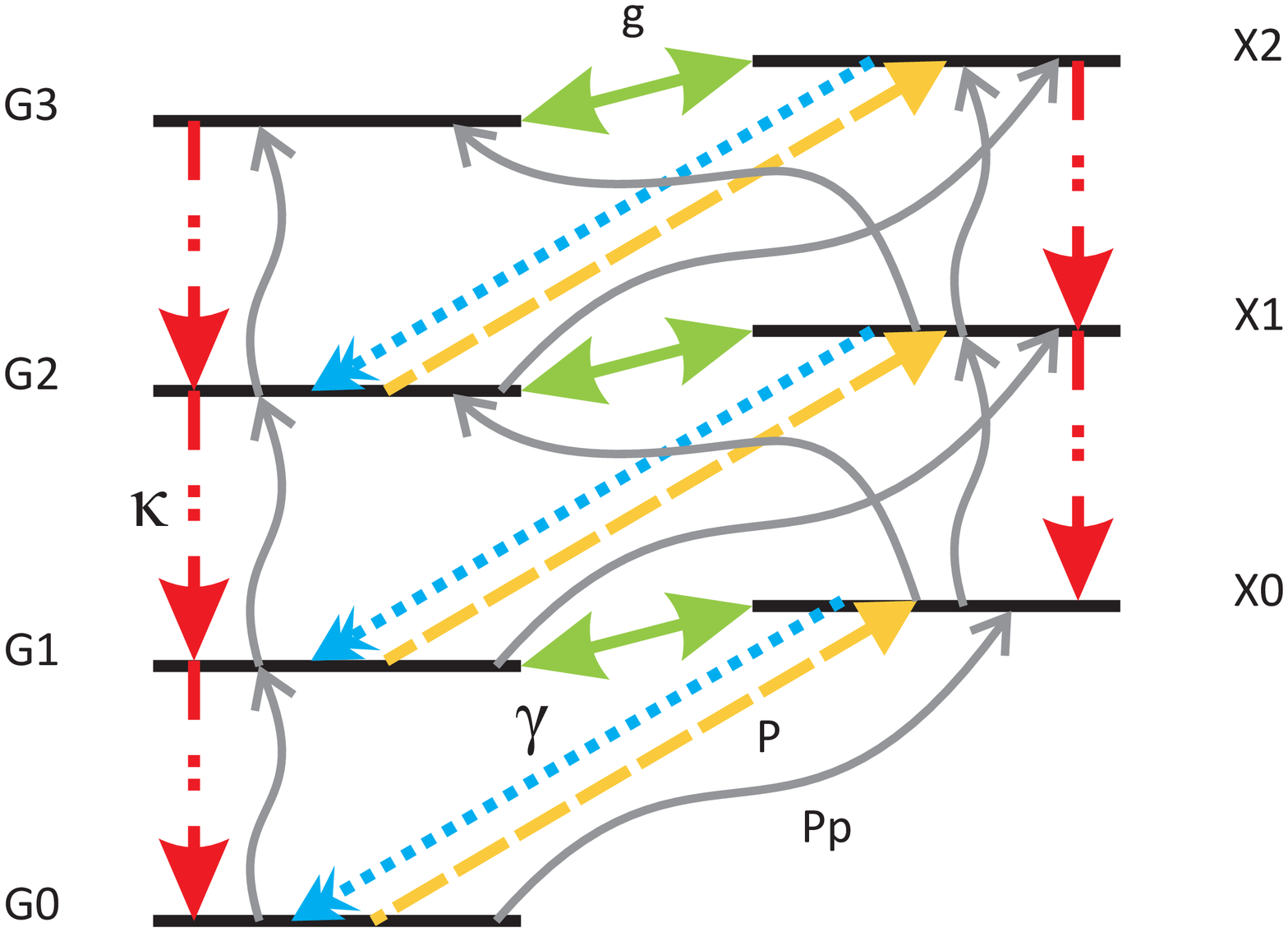}
 \end{minipage}
\ \hfill
 \begin{minipage}{0.48\textwidth}
  \centering b)
\includegraphics[width=1.0\textwidth]{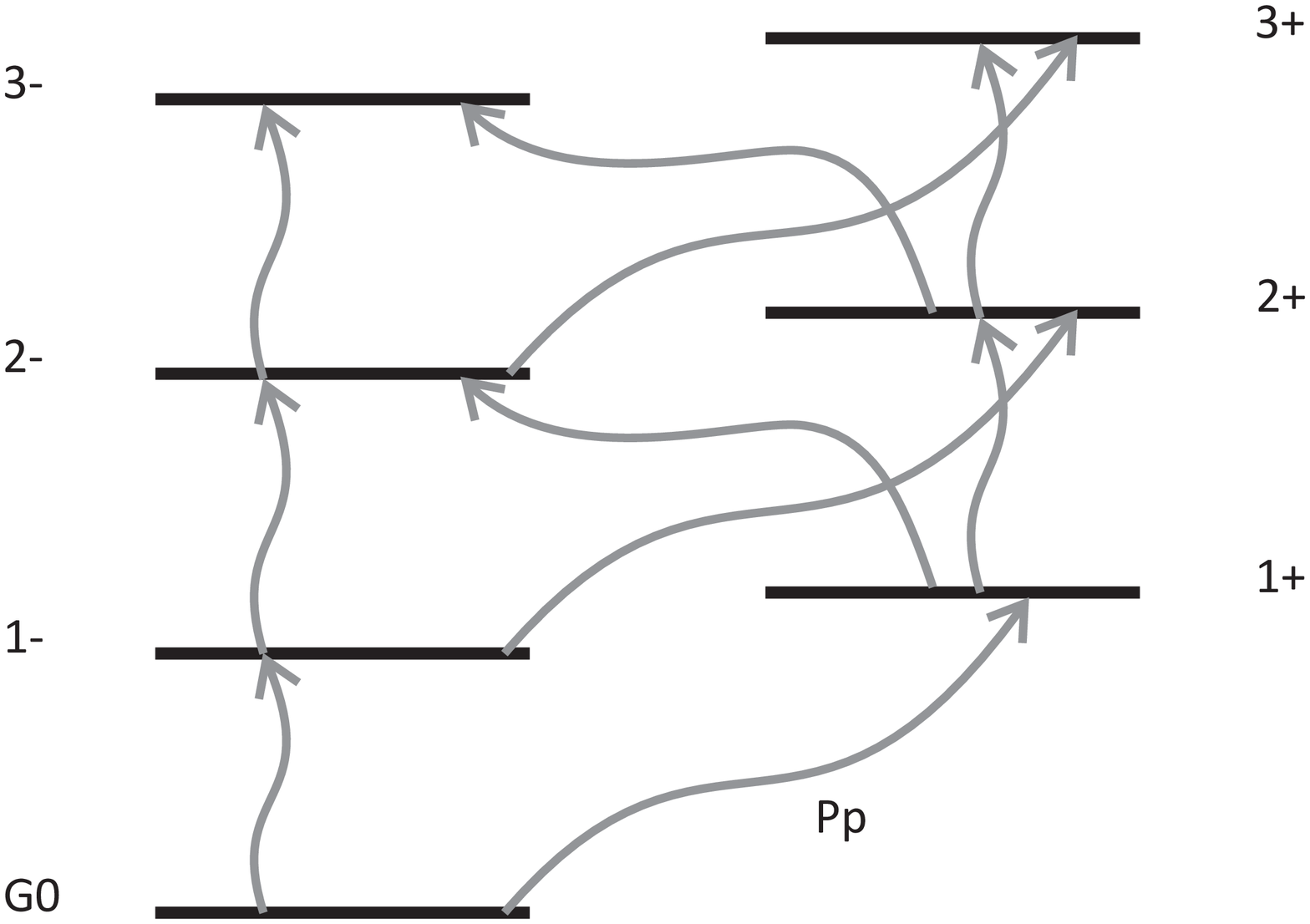}
 \end{minipage}

\caption{a) Ladder of bared states for a two level quantum dot coupled to a
cavity
mode. The double headed green arrow depicts the radiation matter coupling $g$,
dashed yellow lines the exciton pumping rate $P$, dotted blue lines the
spontaneous emission rate $\gamma$, dashed dotted red lines the emission of the
cavity mode $\kappa$ and grey wavy lines the polariton pumping process
$P_p$. b) Transitions due to the polariton pumping term in the master equation 
(\ref{mastereqn}) in the ladder of dressed states. }
\label{transbare}
\label{transdress}
\end{figure}

Since the system we are considering is really an open quantum system the effect
of the environment must be included. The effects of the weak coupling with the
environment are accounted by writing a master equation in the Born-Markov approximation
, for the dynamics of the density operator of the reduced (matter-light) system \cite{petru}.
This master equation accounts for the following processes:
\begin{enumerate}
 \item The continuous and incoherent pumping of the exciton. 
 \item The direct coupling of the exciton to the leaky modes which induces the spontaneous emission process.
 \item The escape of cavity mode photons out of the microcavity due to
incomplete reflectance of the mirrors, the so called coherent emission. These photons are the ones that are measured to obtain the emission spectrum of the system.
 \item An incoherent pumping of polariton (dressed) states.
\end{enumerate}
The first three processes have been discussed in detail in references
\cite{Finley,teje3,Tejedor04,Vera09JPCM} and we shall omit their details here.
An schematic representation of the action of the processes involved in the
dynamics of the systems is given in panel a) of figure \ref{transdress}.
The fourth process of the above list is intended to cause incoherent transitions
among dressed states of two consecutive excitation manifolds $\ket{n,
\pm}\longrightarrow \ket{n+1,\pm}$ as figure
(\ref{transdress}) panel b) shows. This is the equivalent prescription of the
polariton pumping considered in \cite{Vera09,Vera09b}, in which the dressed states of the light
matter Hamiltonian are obtained and are used to define raising and lowering
operators between two consecutive excitation manifolds of polaritonic states.
In complete analogy with \cite{Vera09,Vera09b} we introduce the following lowering operators between polaritonic
states:

\begin{tabular}{l r}
 $P_{++n}=\ket{n,+}\bra{n+1,+}$ & $P_{--n}=\ket{n,-}\bra{n+1,-}$ \\
 $P_{+-n}=\ket{n,-}\bra{n+1,+}$ & $P_{-+n}=\ket{n,+}\bra{n+1,-}$
\end{tabular}

\noindent
With this definitions the master equation for the system takes the form:
\begin{eqnarray}
\label{mastereqn}
 \frac{d}{dt} \rho&=&i [\rho,H]+\frac{\kappa}{2} (2 a \rho
a^{\dagger}-a^\dagger a \rho-\rho a^\dagger a)+\frac{\gamma}{2}(2 \sigma \rho
\sigma^\dagger-\sigma^\dagger\sigma \rho-\rho \sigma^\dagger
\sigma)\\ & &+\frac{P}{2}(2 \sigma^\dagger \rho \sigma-\sigma \sigma^\dagger 
\rho-\rho \sigma \sigma^\dagger) \nonumber \\
& &+\frac{P_p}{2}\sum_{n,i,j}\left[  2
P_{ijn}^{\dagger} \rho P_{ijn}-P_{ijn}
P_{ijn}^{\dagger} \rho- \rho P_{ijn} P_{ijn}^{\dagger} \right] \nonumber.
\end{eqnarray}
\noindent
Where $\kappa$ is the decay rate of the cavity photons due to the incomplete reflectance of the cavity mirrors, $\gamma$ is the exciton decay
rate due to spontaneous emission, $P$ is the rate at which
excitons are being pumped and $P_p$ is the rate at which polaritons are being
pumped. The indices $i$ and $j$ take the values $\{+,-\}$ and
$n\in \mathbb{N}$. The polaritonic bath that is associated to the polariton pumping term in the last 
equation can be thought as a resonant coupling between the bare electron states and the intersubband cavity polariton excitations \cite{Ciuti}.
 
For studying the dynamics of the system we shall write the equations of motion of
the matrix elements taken in the bared basis.  The evaluation of this elements
is straightforward except for the term $L_{P_p}[\rho]=\frac{1}{2}\sum_{n,i,j} \left[ 2
P_{ijn}^{\dagger} \rho P_{ijn}-P_{ijn}
P_{ijn}^{\dagger} \rho- \rho P_{ijn} P_{ijn}^{\dagger} \right]$.
Using the fact that $
\ket{n,+}\bra{n,+}+\ket{n,-}\bra{n,-}=\ket{Gn}\bra{Gn}+
\ket{Xn-1}\bra{Xn-1}
$ and that the trace of an operator is invariant under unitary transformations
it is easily seen that:
\begin{eqnarray}
\bra{i,n} L_{P_p}[\rho] \ket{j,m}&=&\delta _{G,i} \delta
_{G,j} \delta _{m,n} \rho _{Gm-1,Gm-1}+\delta
   _{i,X} \delta _{j,X} \delta _{m,n} \rho _{Gn,Gn} \\
& &+\delta _{G,i} \delta _{G,j} \delta _{m,n} \rho_{Xm-2,Xm-2}+\delta _{i,X}
\delta _{j,X} \delta _{m,n} \rho_{Xn-1,Xn-1} \nonumber\\
& &-\delta _{G,i}\rho _{Gn,jm}-\delta _{i,X} \rho _{Xn,jm}-\delta _{G,j}
\rho_{in,Gm}-\delta _{j,X} \rho _{in,Xm}, \nonumber
\end{eqnarray}
which is independent of both $g$ and $\Delta$. The dynamical equations for the populations and coherences in the bare basis are presented in the appendix \ref{appendix_A}. For simplicity in what follows we have taken as unit of frequency (equivalently
energy) the Rabi constant $g=1$ meV which is a typical value of the light-matter 
coupling constant in semiconductors. All the quantities in what follows are given in units
of $g$.

\subsection{Quantum Regression Theorem and the Emission Spectrum}
One of the few things that can be measured directly from a quantum system is its
spectrum. To obtain the emission spectrum of a system we need 
to take the Fourier transform of the first order correlation function
$\braket{a^{\dagger}(t+\tau)a(t)}$, which requires the knowledge of
the expectation value of two operators at different times. To
 obtain the dynamical equation of such correlation function we take
advantage of the Quantum Regression Theorem (QRT) \cite{walls} which states
that given a set of operators $O_j$ satisfying, $
\frac{d}{d\tau} \braket{O_j(t+\tau)} = \sum_k L_{jk} \braket{O_k(t+\tau)},$
then $\frac{d}{d\tau} \braket{O_j(t+\tau)O(t)} = \sum_k L_{jk}
    \braket{O_k(t+\tau)O(t)}$ for any operator $O$.
\noindent
We follow Tejedor and coworkers \cite{Tejedor04} and write $\langle
a^\dagger(t+\tau) a(t) \rangle =\sum_n \sqrt{n+1} \left(
\braket{a^{\dagger}_{Gn}(t+\tau) a (t)}+\braket{a^{\dagger}_{Xn}(t+\tau) a(t)} \right),$
where the following definitions have been used:
\begin{eqnarray}
\label{varsdef}
  &&a^\dag_{Gn} = \ket{Gn+1}\bra{Gn} \\
  &&a^\dag_{Xn} = \ket{Xn+1}\bra{Xn} \nonumber \\
  &&\sigma^\dag_{n} = \ket{Xn}\bra{Gn} \nonumber \\
  &&\zeta _n = \ket{Gn+1}\bra{Xn-1}\nonumber.
\end{eqnarray}
Note that these operators act between two consecutive excitation manifolds. It turns out that the last set operators satisfy the following set of closed
differential equations:
\begin{eqnarray}
\label{qrt}
\frac{d} {d\tau} \langle a^\dagger_{Gn} (\tau) \rangle &=& \left(-P-2 P_p-i
\Delta -n
\kappa-\frac{\kappa }{2}+i \omega_X \right) \langle
 a^\dagger_{Gn}(\tau) \rangle\\ & &
+\kappa  \sqrt{(n+1)(n+2)} \langle
a^\dagger_{Gn+1}(\tau ) \rangle
\nonumber \\
& &+\gamma \langle a^\dagger_{Xn}(\tau ) \rangle-i g \sqrt{n} \langle
\zeta_n(\tau ) \rangle +i g
\sqrt{n+1} \langle \sigma^\dagger_n(\tau ) \rangle
\nonumber \\
\frac{d }{d\tau}\langle \sigma^\dagger_{n} (\tau) \rangle &=& i g \sqrt{n+1}
\langle
a^\dagger_{Gn}(\tau ) \rangle-i g   \sqrt{n} \langle a^\dagger_{Xn-1}(\tau )
\rangle  \nonumber\\
& &+\frac{1}{2}
(-P-4P_p-\gamma   -2 n \kappa +2 i \omega_X) \langle \sigma^\dagger_n(\tau   )
\rangle
\nonumber \\
& &+(n+1) \kappa  \langle \sigma^\dagger_{n+1}(\tau ) \rangle
\nonumber \\
\frac{d }{d\tau} \langle a^\dagger_{Xn-1} (\tau) \rangle&=& P \langle
a^\dagger_{Gn-1}(\tau )
\rangle +\kappa \sqrt{n(n+1)} \langle a^\dagger_{Xn}(\tau
) \rangle \nonumber \\
& & +\frac{1}{2} (-4P_p-2 \gamma -2 i \Delta -2 n\kappa +\kappa +2 i
\omega_X) \langle
a^\dagger_{Xn-1}(\tau ) \rangle
\nonumber \\
& &+i g \sqrt{n+1} \langle \zeta_n(\tau ) \rangle-i g \sqrt{n} \langle
\sigma^\dagger_n(\tau )
\rangle
\nonumber \\
\frac{d }{d\tau}\langle \zeta_n (\tau)\rangle&=&-i g \sqrt{n} \langle
a^\dagger_{Gn}(\tau )
\rangle+i g \sqrt{n+1}\langle a^\dagger_{Xn-1}(\tau )\rangle \nonumber\\
&&+\frac{1}{2}
(-P-4 P_p-\gamma -4 i
\Delta -2 n \kappa +2 i \omega_X) \langle \zeta_n(\tau ) \rangle  \nonumber \\
\nonumber
& &+\sqrt{n(n+2)} \kappa   \langle \zeta_{n+1}(\tau ) \rangle. \nonumber
\end{eqnarray}
The QRT implies that the two time operators
$\braket{a^\dag_{Gn}(t+\tau) a(t)}$, $\braket{a^\dag_{Xn}(t+\tau) a(t)}$, $
\braket{\sigma^\dag_{n}(t+\tau) a(t)}$, $\braket{\zeta _n(t+\tau) a(t)}$
satisfy equations (\ref{qrt}), subject to the initial conditions:
\begin{eqnarray}
\label{qrtinit}
\braket{a^\dag_{Gn}(t) a(t)}&=& \sqrt{n+1} \rho_{Gn+1,Gn+1}(t) \\
\braket{a^\dag_{Xn}(t) a(t)}&=& \sqrt{n+1} \rho_{Xn+1,Xn+1}(t) \nonumber \\
\braket{\sigma^\dag_{n}(t) a(t)} &=& \sqrt{n+1}\rho_{Gn+1,Xn}(t) \nonumber \\
\braket{\zeta _n(t) a(t)} &=& \sqrt{n} \rho_{Xn,Gn+1}(t). \nonumber
\end{eqnarray}
The role of the parameters $\omega_X, \Delta, g,
\kappa,\gamma,P,P_p$ is twofold, on the one hand they determine the dynamics of
the two time operators via (\ref{qrt}), and, on the other hand they set the
initial conditions (\ref{qrtinit}) that shall be propagated according to the
dynamical equations (\ref{qrt}). Here we will be interested in the light that
the systems emits in the stationary limit so that the limit $t\rightarrow
\infty$ will be taken in equation (\ref{qrtinit}).

\subsection{Entanglement \label{epp}}
To quantify the entanglement between excitonic and photonic subsystems we use the
Peres criterion \cite{peres}. This criterion says that if the state of bipartite
system is separable then the eigenvalues of its partial transpose are all
positive. For our case the density matrices, $\rho$, we are considering have
only the following non-zero matrix elements:
$\rho_{Xn,Xn}, \rho_{Gn+1,Gn+1}, \rho_{Gn+1,Xn}, \rho_{Xn,Gn+1}$ correspondingly
the nonzero elements of the partial transpose $\rho^{\Gamma}$ respect to
excitonic subsystem are:
\begin{eqnarray}
\rho^{\Gamma}_{Gn,Gn}&=&\rho_{Gn,Gn} \\
\rho^{\Gamma}_{Gn+1,Gn+1}&=&\rho_{Gn+1,Gn+1} \nonumber\\
\rho^{\Gamma}_{Xn+1,Gn}&=&\rho_{Gn+1,Xn} \nonumber\\
\rho^{\Gamma}_{Gn,Xn+1}&=&\rho_{Xn,Gn+1} \nonumber
\end{eqnarray}
The matrix $\rho^{\Gamma}$ takes a block diagonal form of blocks
$1\times1$ and $2\times2$, when written in the basis
$\{ \ket{X0},\ket{G0},\ket{X1},\ket{G1},\ket{X2} \ldots
\ket{Gm},\ket{Xm+1},\ket{Gm+1} \}$ (Notice the ordering of the basis):
\begin{equation}
\rho^{\Gamma}=\left(
\begin{array}{lllllll}
 \rho _{X0,X0} &    &  &  & &   &\\
  &              \ddots &  & & & &\\
  &                      & \rho _{Gn,Gn} & \rho_{Xn,Gn+1} & & &\\
  & &   \rho_{Gn+1,Xn}& \rho _{Xn+1,Xn+1} & & &\\
  &               &  & & \ddots & &\\
  & &                 &                   & & &\rho_{Gm+1,Gm+1}
\end{array}
\right).
\end{equation}
The eigenvalues of the last matrix are easily obtained. The ones corresponding to the left upper and right lower entries (blocks)
are $\rho_{X0,X0}$ and $\rho_{Gm+1,Gm+1}$ and are always positive ore zero. 
The ones corresponding to the 2 $\times$ 2 blocks are $\frac{1}{2}\left(\rho _{Gm,Gm}+\rho _{Xm+1,Xm+1} \pm \sqrt{(\rho_{Gm,Gm}-\rho_{Xm+1,Xm+1})^2+4 |\rho_{Xm,Gm+1}|^2}\right) $. 
%\begin{eqnarray}
%\lambda(\rho^{\Gamma})=& &\{\rho_{X0,X0}, \rho_{Gm+1,Gm+1},\ldots \\
%  & &\frac{1}{2}(\rho _{Gm,Gm}+\rho _{Xm+1,Xm+1} \pm \nonumber \\ & &\sqrt{(\rho_{Gm,Gm}-\rho_{Xm+1,Xm+1})^2+4 |\rho_{Xm,Gm+1}|^2}) \}. \nonumber
%\end{eqnarray}
In order to have a negative eigenvalue (and an entangled state) the following
condition must be met for some $n$:
\begin{equation}
 |\rho_{Xn,Gn+1}|>\sqrt{\rho_{Gn,Gn}\rho_{Xn+1,Xn+1}}.
\end{equation}
Notice that the above inequality can also be obtained by using the criterion recently derived in \cite{asp}. 
Then, we can quantify entanglement using the following function, which is
equivalent to the Peres criterion:
\begin{equation}
\label{ep}
 E(\rho)= 4 \sum_n \left(\max \left\{0,
|\rho_{Xn,Gn+1}|-\sqrt{\rho_{Gn,Gn}\rho_{Xn+1,Xn+1}} \right\}\right)^2.
\end{equation}
For Bell-like states $\rho_{Bell}=\ket{\psi} \bra{\psi}$,
$\ket{\psi}=\frac{1}{\sqrt{2}}\left(\ket{Gn+1}+
e^{i \phi} \ket{X n}\right)$, $\phi \in \mathbb{R}$, $E(\rho)$ will equal 1. In,
particular the polaritonic states $\ket{n, \pm}$ in resonance have $E(\rho)=1$.

\section{Results and Discussion}

\subsection{Exciton Pumping vs. Polariton Pumping}
\begin{figure}[ht]
  \begin{minipage}{0.3\textwidth}    
\centering
$\braket{n}$
\bigskip
  \end{minipage}
  \ \hfill
  \begin{minipage}{0.3\textwidth}    
\centering
$g^2(\tau=0)$
\bigskip
  \end{minipage}
  \ \hfill
  \begin{minipage}{0.3\textwidth}    
\centering
$\braket{\sigma_z}$
\bigskip
  \end{minipage}
\\
  \begin{minipage}{0.3\textwidth}    
    \includegraphics[width=\textwidth]{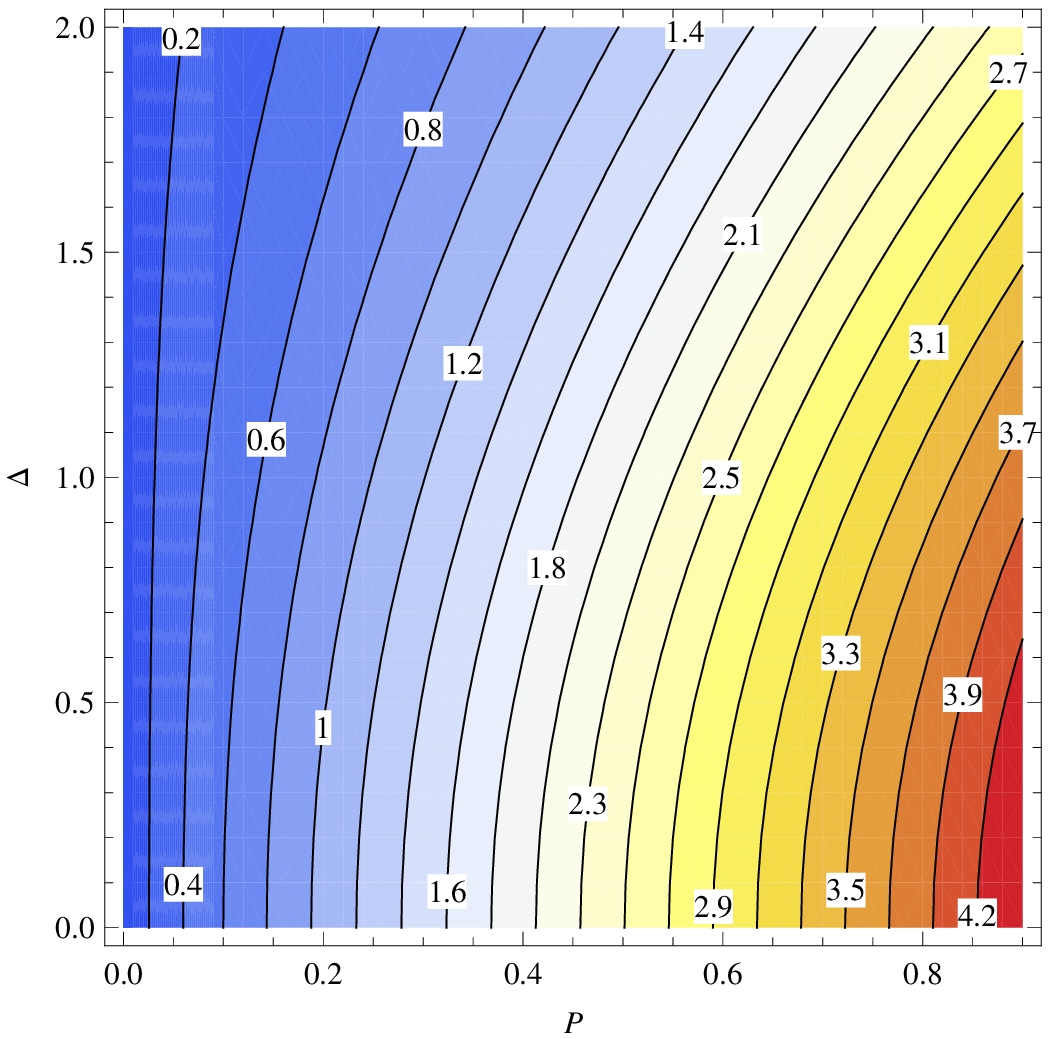}
  \end{minipage}
  \ \hfill
  \begin{minipage}{0.3\textwidth}    
    \includegraphics[width=\textwidth]{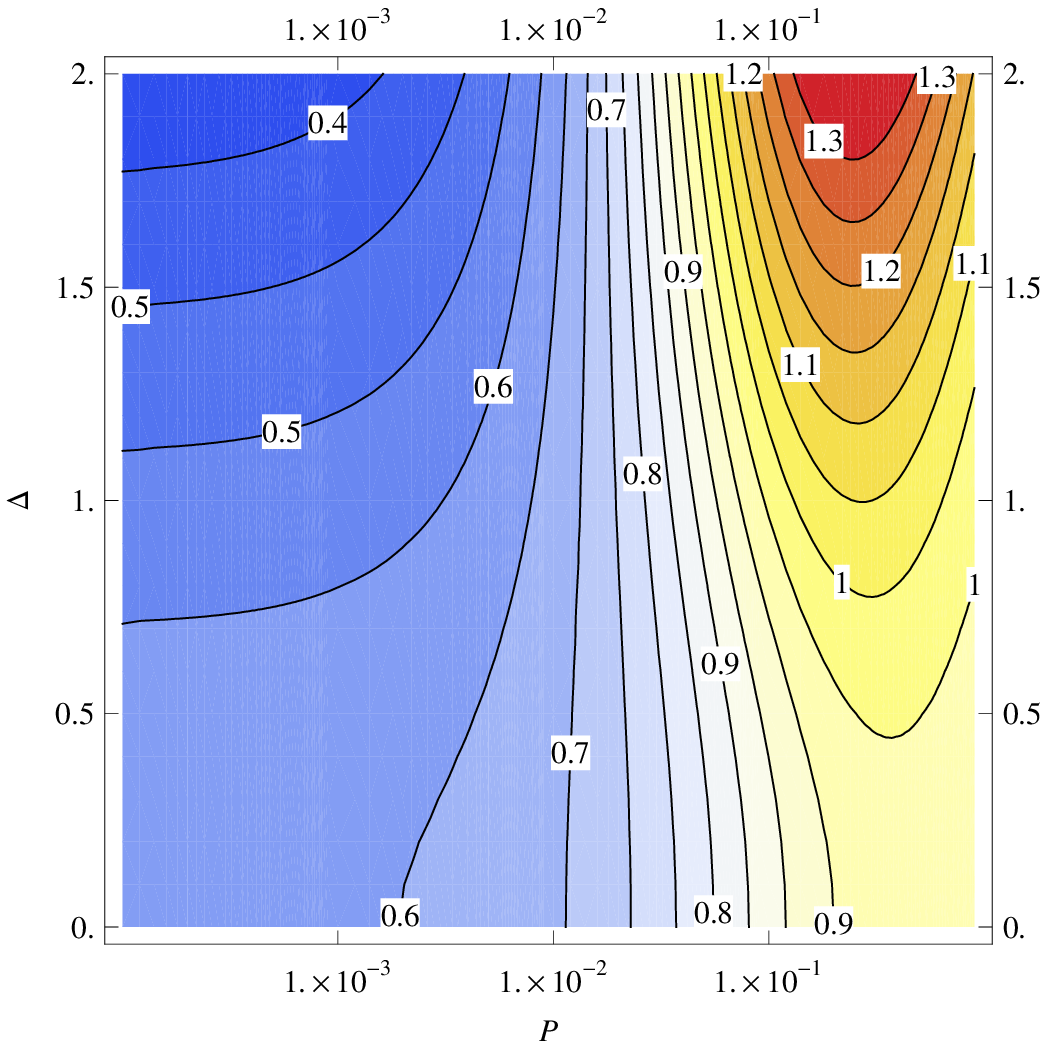}
  \end{minipage}
  \ \hfill
  \begin{minipage}{0.3\textwidth}    
    \includegraphics[width=\textwidth]{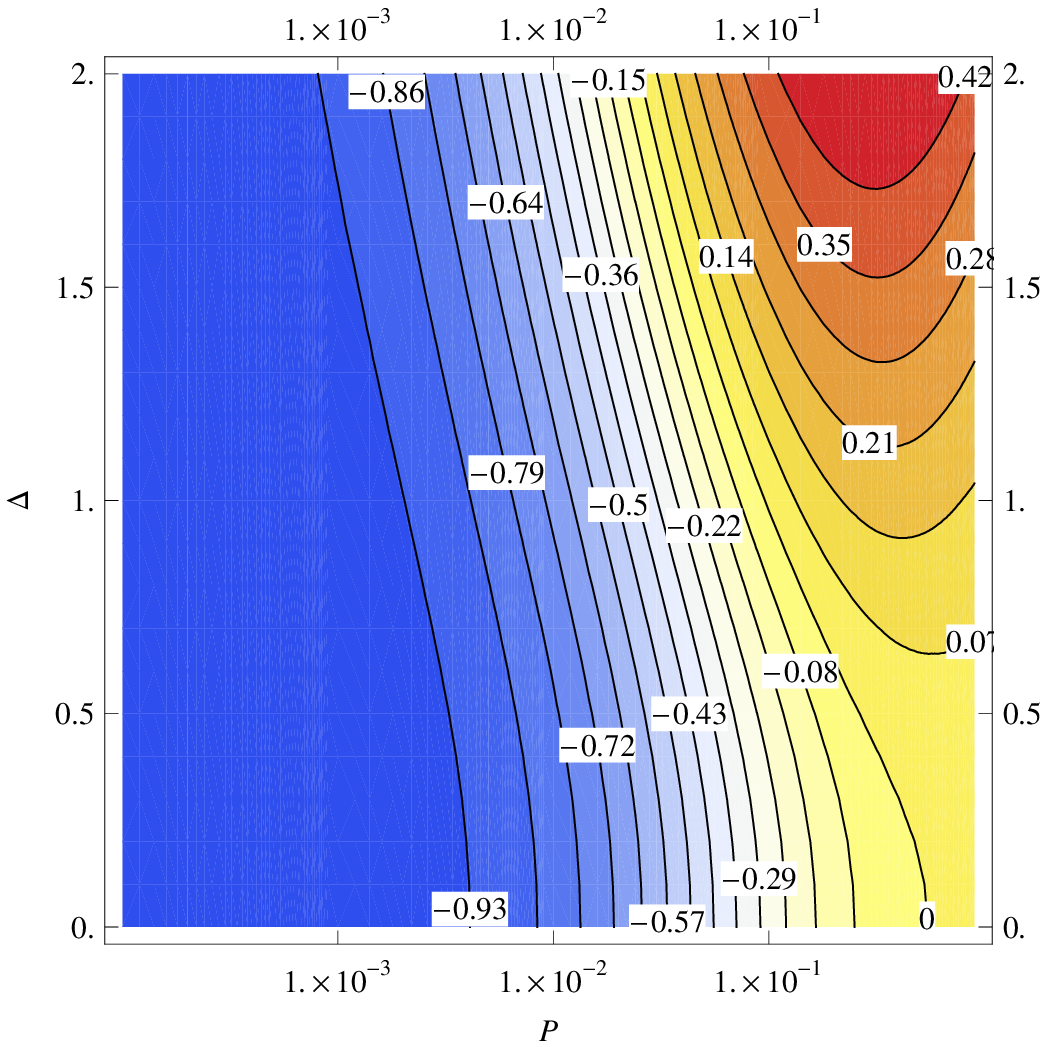}
  \end{minipage}
\\
  \begin{minipage}{0.3\textwidth}    
    \includegraphics[width=\textwidth]{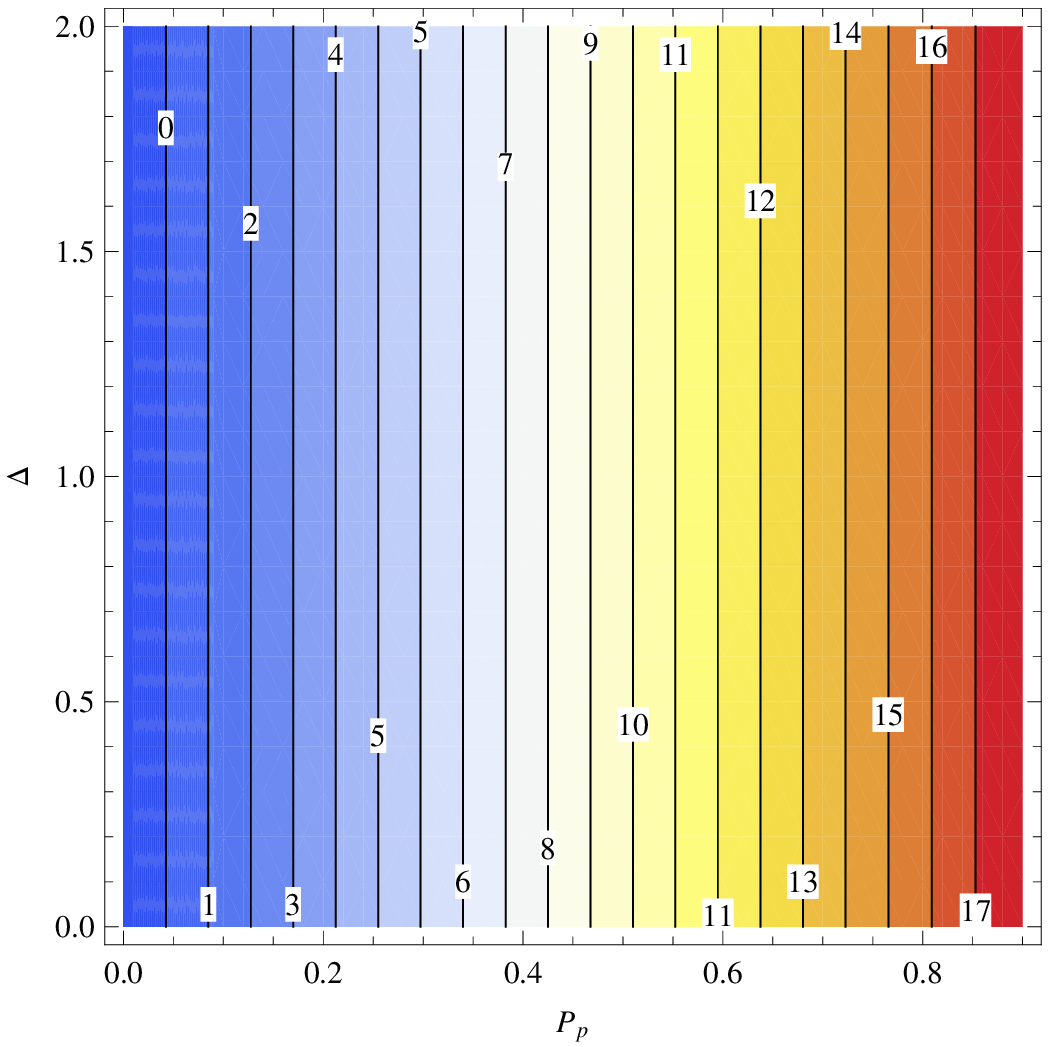}
  \end{minipage}
\ \hfill
  \begin{minipage}{0.3\textwidth}    
    \includegraphics[width=\textwidth]{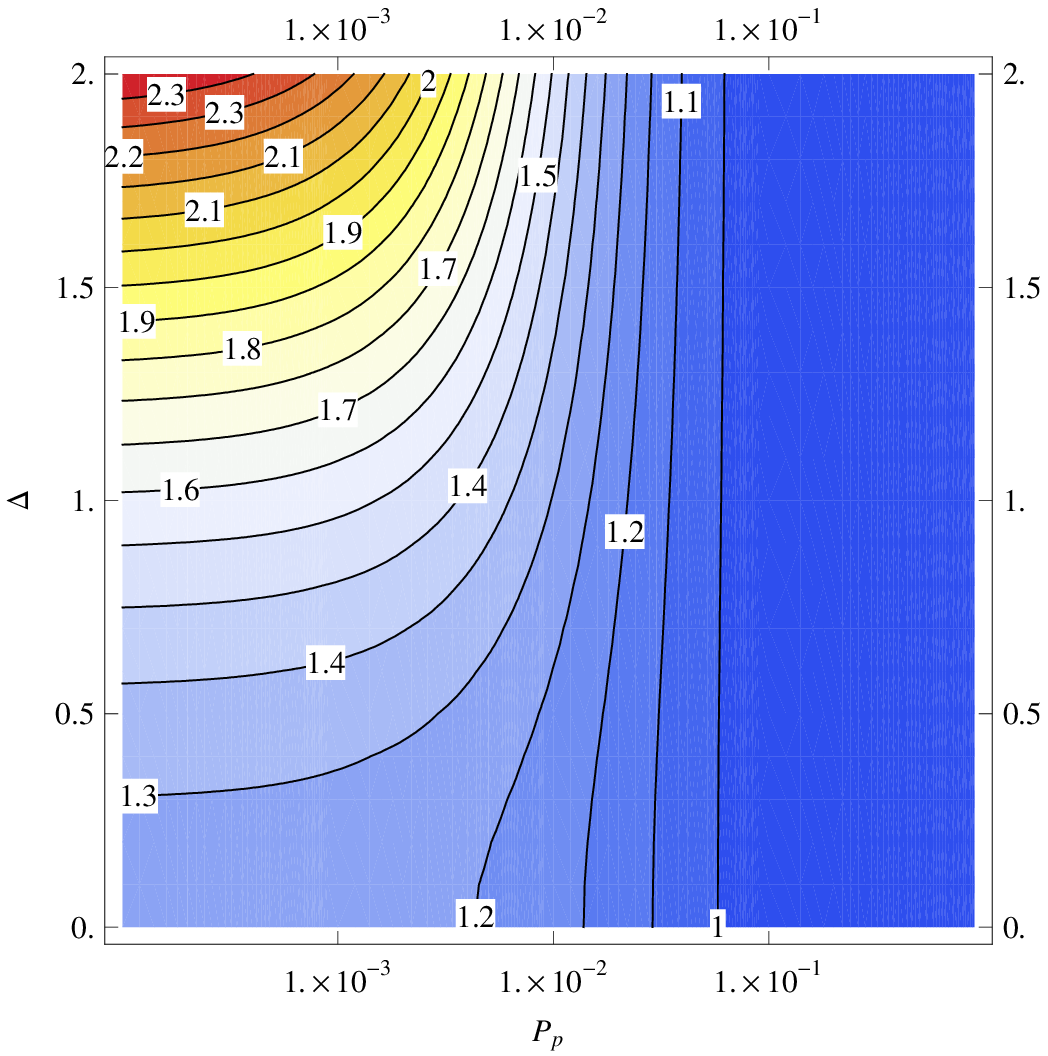}
  \end{minipage}
  \ \hfill
  \begin{minipage}{0.3\textwidth}    
    \includegraphics[width=\textwidth]{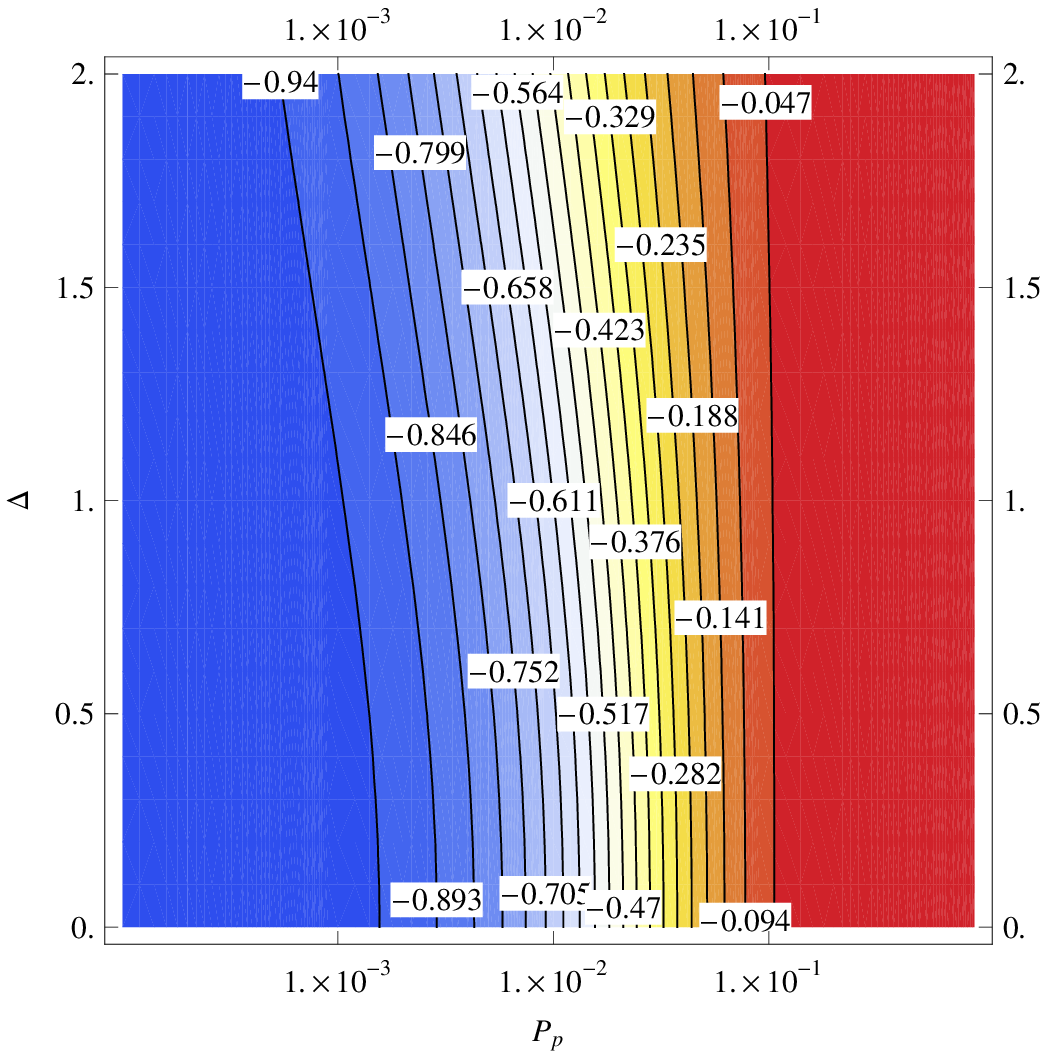}
  \end{minipage}
  \caption{\label{nphotons}Average number of photons, second order coherence
function at zero delay and population inversion  as a function of the detuning
$\Delta$ and
the exciton pumping rate $P$ or the polariton
pumping $P_p$. Parameters: $\kappa=0.1$, $g=1$, $\omega_X=1000$, $\gamma=0$, for
the upper panel $P_p=0$ and for the lower panel $P=0$. In most experimental
situations $\gamma$ is at least two orders of magnitude smaller than the rest of
the parameters \cite{yama}. For this calculations the Fock space was truncated
in $n_{\max}=40$}.
 \end{figure}

In this section, we compare the evolution of some observables in the stationary
limit as a function of the detuning ($\Delta$) and the pumping
rates of both excitons ($P$) and polaritons ($P_p$).
In Figure \ref{nphotons}, we present the evolution of the average number of
photons, $\braket{n}=\braket{a^{\dagger}a}$ in the stationary limit for the case
of strong coupling ($|\kappa-\gamma|/4 \ll g$) as a function of $\Delta$ and
either $P$ or $P_p$.
From this figure several conclusions can be drawn. First the pumping of
polaritons is more efficient accumulating photons in the cavity, this
can be understood if one sees the type of transitions that both types of
pumpings cause in the bare basis ladder of states as it is seen in Figure 
\ref{transbare}. On the one hand $P$ causes
diagonal transitions, whereas $P_p$ induces vertical (and diagonal) transitions that can
rapidly populate states with a high number of photons. Secondly, and rather
surprisingly, the average number of photons (and as we shall comment later)
the observables we have monitored are not sensible to the detuning when the
term $P_p>\kappa=0.1$.\\

We have also calculated the effect of both mechanisms on the population
inversion $\braket{\sigma_z}$ and the second order coherence function at zero
delay $g^2(\tau=0)=\braket{a^{\dagger} a^{\dagger} a
a}/\braket{a^{\dagger}a}^2$. 
In Figure \ref{nphotons} we present the contour plots of $g^2(\tau=0)$ and
$\braket{\sigma_z}$. 
It is clearly seen that one can explore different field
statistics as $P$ and $\Delta$ are varied and that the population inversion
grows as a function of $P$. The corresponding results for the polariton
pumping mechanism can be summarized as follows:
\begin{itemize}
 \item  The second order coherence function is almost equal to one,
 and is independent of $\Delta$ for values of $P_p$ larger than
$\kappa=0.1$. 
 \item The population inversion presents a similar behavior to $g^2(\tau=0)$.
It approaches to 0 from below as $P_p$ is increased  and it is nearly
independent of $\Delta$ for values of $P_p>\kappa=0.1$
\end{itemize}
Comparing the results of both types of pumping it is seen that the polariton
pumping mechanism is unable to cause a positive population inversion, can
greatly increase the intensity of the light stored in the cavity, and the light
stored in the cavity might have Poisson statistics, since the variance of the
number of photons equals its mean value (although in general it will not be a
coherent state since the reduced density matrix of the photons will be a mixed state).
The first observation is understandable
when one inspects the action of the term $P_p$ in the bare basis. On the one
hand the term $P_p$ causes transitions with equal intensity from
$\ket{X}\longrightarrow\ket{G}$ and from $\ket{G}\longrightarrow\ket{X}$, this
explains why $\braket{\sigma_z}$ is close to zero. The fact of being always
slightly negative is related to the asymmetry that the vacuum state
$\ket{G0}$
introduces between the states $G$ and $X$, in such a way that state vectors with
no exciton have a slightly heavier statistical weight. The second observation is
related to the possibility of having an inversionless polaritonic laser \cite{Vera09b,BlochJ}.

\subsection{Emission Spectrum}
Since the effects of the excitonic pumping have been
extensively studied in \cite{teje1,teje2,teje3} in this section we only study
the effects of the polaritonic pumping in the emission spectrum of the system.
To do so we write the equations of motion of the variables in
equation (\ref{qrt}) as:
\begin{equation}
\label{eqn:dyn}
 \frac{d}{dt} \mathbf{v}(t)= \mathbf{A}(P_p,P,g,\kappa,\gamma,\omega_X,\Delta) \mathbf{v}(t)
\end{equation}
 where $ \mathbf{v}(t)=\{ \braket{\sigma^\dag_{0}(t)}, \braket{a^\dag_{G0}(t)}, 
\ldots,  \braket{\zeta
_n(t)}, \braket{\sigma^\dag_{n}(t)}, \braket{a^\dag_{Xn-1}(t)},
\braket{a^\dag_{Gn}(t)},\ldots \}$. The formal solution of equation
(\ref{eqn:dyn}) is given by:
\begin{equation}
\label{eqn:formalsol}
 \mathbf{v}(t+\tau)=\exp(\mathbf{A}(P_p,P,g,\kappa,\gamma,\omega_X,\Delta) \tau) \mathbf{v}(t)
\end{equation}
From equations (\ref{qrt}) one sees that $
\mathbf{A}(P_p,P,g,\kappa,\gamma,\omega_X,\Delta) = -2
P_p \mathbf{I}$ $+\mathbf{B}(P,g,\kappa,\gamma,\omega_X,\Delta) $ ($\mathbf{I}$ is the identity matrix),
\emph{i.e.} the polariton pumping term  is diagonal in equation (\ref{qrt}),
and because it commutes with the matrix $\mathbf{B}(P,g,\kappa,\gamma,\omega_X,\Delta)$
it can be factored out in equation (\ref{eqn:formalsol}) as follows 
$\mathbf{v}(t+\tau)=\exp(- 2 P_p \tau ) \exp(\mathbf{B}(P,g,\kappa,\gamma,\omega_X,\Delta) \tau) \mathbf{v}(t)$

The last equation implies that the pumping rate $P_p$ \emph{cannot} modify the
oscillation frequencies of the first order correlation function since it acts as
a common overall decay rate for all the operators involved in equation
(\ref{qrt}). It can only redistribute the statistical weights of the different
frequencies by modifying the initial values of the two time
operators that are related to the populations and coherences in
equation (\ref{qrtinit}).
For instance in the case where one considers transitions between
the ground state $\ket{G0}$ and the states $\ket{X0}$ and $\ket{G1}$ only two
operators appear in the equations of the QRT,
$\braket{a^{\dagger}_{G0}(t)}$ and $\braket{\sigma^{\dagger}_{G0}(t)}$. This
approximation is valid when the pumping ($P$ or $P_p$) is small
enough as compared with the losses ($\gamma$ and $\kappa$) to
not have an average photon number of more than one \cite{Vera09JPCM}. In
this case the equations of the QRT are:
\begin{eqnarray}
& &
\frac{d}{dt}
 \left(
\begin{array}{l}
 \braket{a^{\dagger}_{G0}(t)}\\
 \braket{\sigma^{\dagger}_{G0}(t)}
\end{array}
\right)
=\\
& &\left(
\begin{array}{ll}
  -\frac{P}{2}-2 P_p-\frac{\gamma}{2} + i \omega_X & i g \\
 i g &- P-2 P_p- i \Delta - \frac{\kappa}{2} + i \omega_X
\end{array}
\right)
 \left(
\begin{array}{l}
 \braket{a^{\dagger}_{G0}(t)}\\
 \braket{\sigma^{\dagger}_{G0}(t)}
\end{array}
\right)  \nonumber
\end{eqnarray}
The eigenvalues $\lambda_{\pm}$ of the square matrix in the last equation will
be related to the positions $\omega_{\pm}$ and widths $\Gamma_{\pm}$ of the
emission spectrum ($i \lambda_{\pm}=\omega_{\pm}+i \Gamma_{\pm}$) and are given
by:
\begin{eqnarray}
\label{2qrt}
 \lambda_{\pm}=& & \frac{1}{4} (-3 P-8 P_p-\gamma -2 i \Delta -\kappa +4 i
   \omega_X \\ & &\pm i \sqrt{16 g^2-(P-\gamma +2 i \Delta +\kappa )^2})
\nonumber
\end{eqnarray}
As expected the term $P_p$ only enters as a decay rate that widens both peaks of
emission in the same way.\\
Actually one can go a bit further in the analytical calculation by considering
the case where there is no exciton pumping, $P=0$. In this case one can obtain
analytically \emph{all} the eigenvalues of the matrix
$\mathbf{A}=\mathbf{A}(P_P,P=0,g,\kappa,\gamma,\omega_X,\Delta)$. To this end notice that the only
term that couples the group of operators $\braket{\zeta
_n(t)}$, $\braket{\sigma^\dag_{n}(t)}$, $\braket{a^\dag_{Xn-1}(t)}$,
$\braket{a^\dag_{Gn}(t)}$ and $\braket{\zeta
_{n-1}(t)}$, $\braket{\sigma^\dag_{n-1}(t)}$, $\braket{a^\dag_{Xn-2}(t)}$,
$\braket{a^\dag_{Gn-1}(t)}$ in equation (\ref{qrt}) is the term $P
\braket{a^{\dagger}_{Gn-1}}$ in the equation for $\frac{d}{d\tau}
\braket{a^{\dagger}_{Xn-1}} $, so if $P=0$ the structure of the matrix $A$ will
consist of block diagonal terms of sizes $2\times2$ and $4\times4$ and off
diagonal terms \emph{over} the diagonal of the matrix. This implies that
to reduce the matrix $A$ to an upper triangular matrix it is necessary to rotate
each diagonal block and the blocks above it. Once the matrix has an upper
triangular form their eigenvalues are simply given by the elements of the
diagonal. Summarizing, the eigenvalues of the whole matrix $\mathbf{A}$ are simply the
eigenvalues of the blocks $2\times2$ and $4\times4$.\\
The eigenvalues of the $2\times2$ matrix are given by equation (\ref{2qrt}) with $P=0$.
The structure of the blocks $4\times4$ is:
\begin{eqnarray}
\mathbf{A}& &|_{4\times4} =\\
& &
 \left(
\begin{array}{llll}
 i( \omega_X-2 \Delta)  & 0 & i g
   \sqrt{n+1} & -i g \sqrt{n} \nonumber \\
 \,-n \kappa  -\frac{\gamma }{2}-2 P_p  & & & \nonumber\\
 0 & i \omega_X-2 P_p  & -i g \sqrt{n} & i g
   \sqrt{n+1} \nonumber \\
 & \, -\frac{\gamma }{2}-n \kappa & & \nonumber \\
 i g \sqrt{n+1} & -i g \sqrt{n} & i (\omega_X-\Delta) -2 P_p   & 0
\nonumber\\
 & & \, -(n-\frac{1}{2}) \kappa -\gamma  & \nonumber\\
 -i g \sqrt{n} & i g \sqrt{n+1} & 0 & i( \omega_X -i \Delta) \nonumber\\
& & & \,-(n+\frac{1}{2}) \kappa -2P_p \nonumber
\end{array}
\right)
\end{eqnarray}
The eigenvalues of the above matrix are simply given by:
\begin{eqnarray}
 \lambda_{\pm,\pm}&&=-2 P_p -\frac{\gamma }{2}-i \Delta -n \kappa
+i\omega_X \pm \frac{1}{2 \sqrt{2}} \sqrt{a \pm \sqrt{b}}
\end{eqnarray}
\begin{eqnarray}
a =&&  -8 (2 n+1) g^2-4 \Delta ^2+\left(\kappa - \gamma \right)^2 
\nonumber\\
b=&& 256 n (n+1) g^4+\left[ \left(\kappa - \gamma \right)^2 +4 \Delta ^2
 \right]^2  \nonumber\\ & &-16 \left[(2 n+1) (\gamma ^2+\kappa^2+4
\Delta^2)-2 (2 n
\kappa +\kappa -2 i \Delta )
\gamma  -4 i \Delta 
   \kappa \right] g^2 \nonumber
\end{eqnarray}
An interesting question that may be asked is under what conditions the imaginary
parts of the eigenvalues differ at resonance ($\Delta=0$), \emph{i.e.}, under
what condition $a\pm\sqrt{b}<0|_{\Delta=0}$. The condition is simply given by:
\begin{equation}
 |g|>\frac{|\kappa-\gamma|}{4\sqrt{n}}
\end{equation}
Which for $n=1$ reduces to well the known relation for strong coupling in the absence
of exciton pumping \cite{Gerry}, but actually also tells that different
excitation manifolds might be in different regimes so while some excited
manifold might be ``dressed'' while lower states might be ``bared'' \cite{teje3}.
In our case this condition will not be affected by the polariton pumping $P_p$.

\begin{figure}
  \begin{minipage}{0.48\textwidth}
\centering
$P_p=10^{-4}$
 \end{minipage}
\ \hfill
 \begin{minipage}{0.48\textwidth}
\centering
$P_p=10^{-3}$
 \end{minipage}
\\
 \begin{minipage}{0.48\textwidth}
\includegraphics[width=1.0\textwidth]{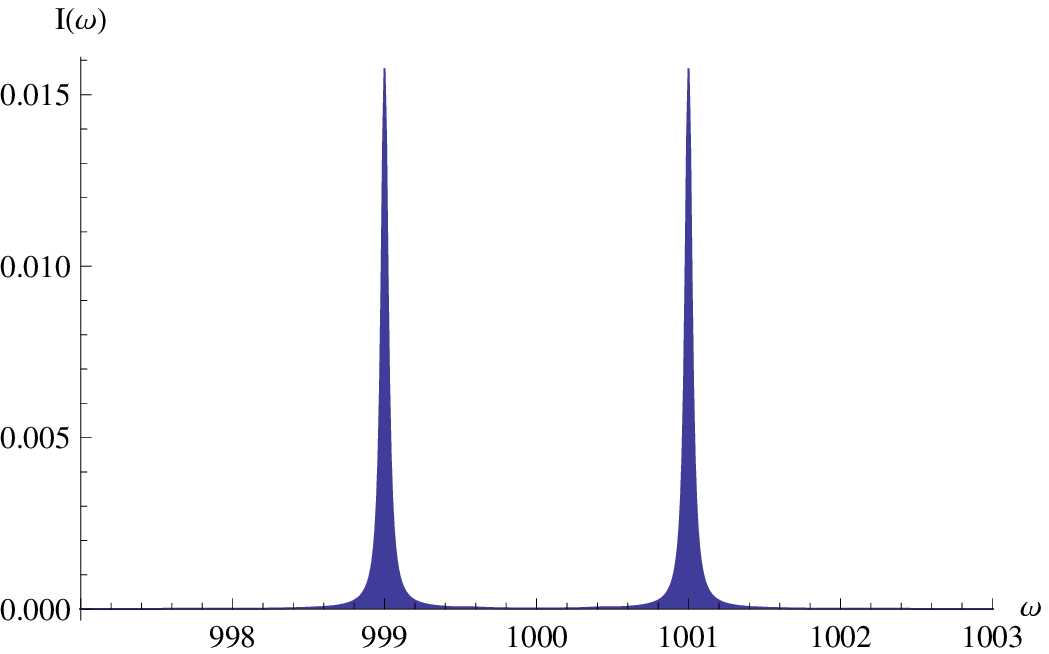}%
 \end{minipage}
\ \hfill
 \begin{minipage}{0.48\textwidth}
\includegraphics[width=1.0\textwidth]{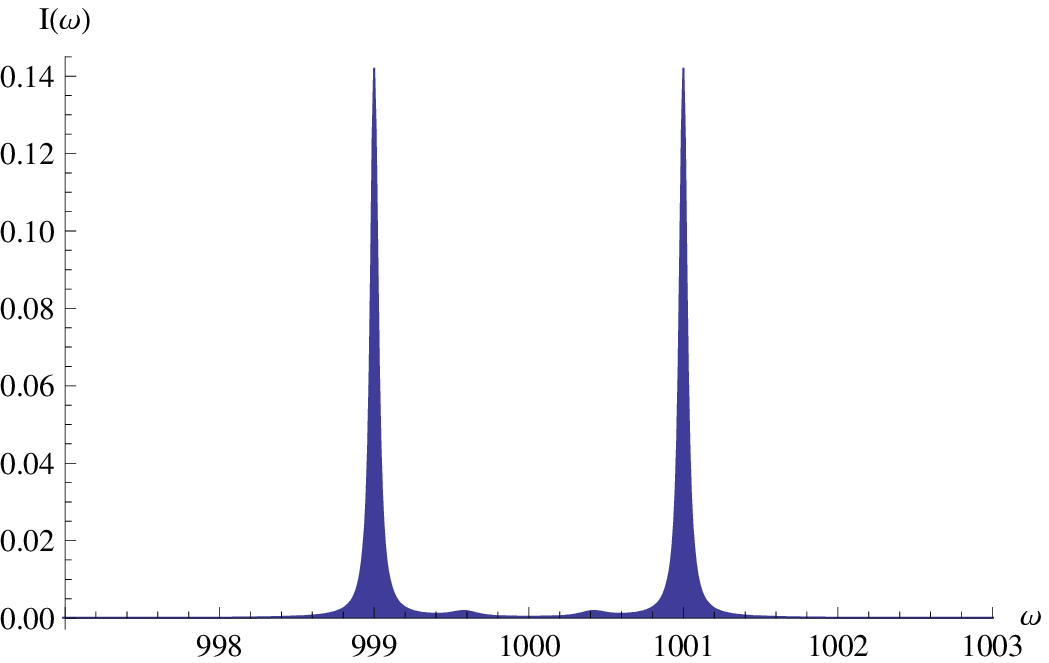}%
 \end{minipage}
\\
  \begin{minipage}{0.48\textwidth}
\centering
\bigskip
$P_p=10^{-2}$
 \end{minipage}
\ \hfill
 \begin{minipage}{0.48\textwidth}
\centering
\bigskip
$P_p=10^{-1}$
 \end{minipage}
\\
 \begin{minipage}{0.48\textwidth}
\includegraphics[width=1.0\textwidth]{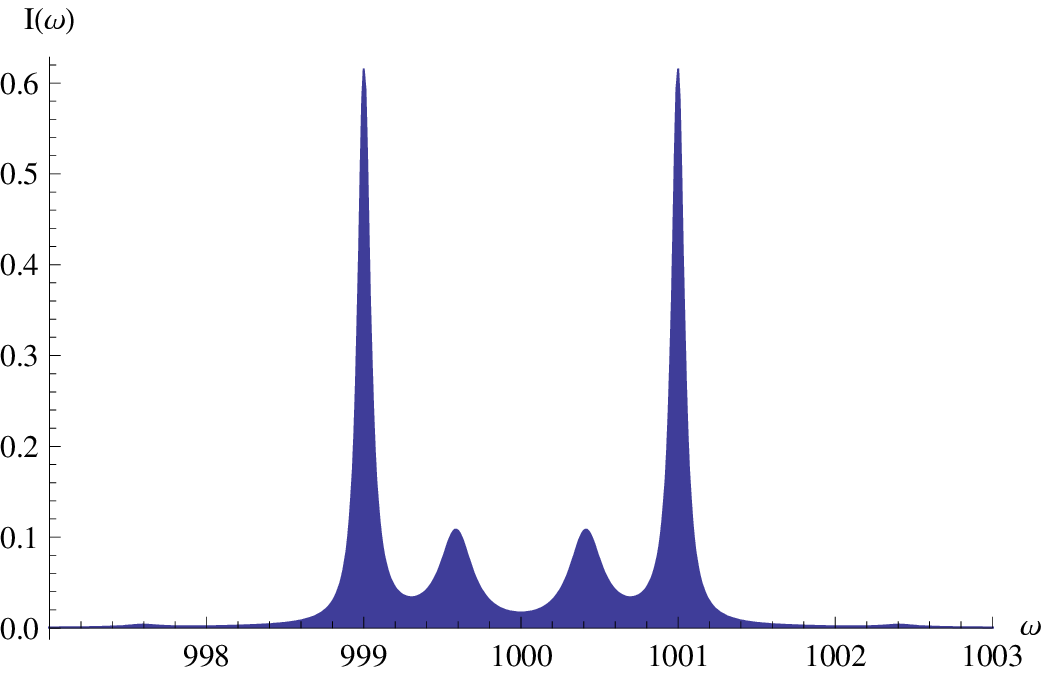}%
 \end{minipage}
\ \hfill
 \begin{minipage}{0.48\textwidth}
\includegraphics[width=1.0\textwidth]{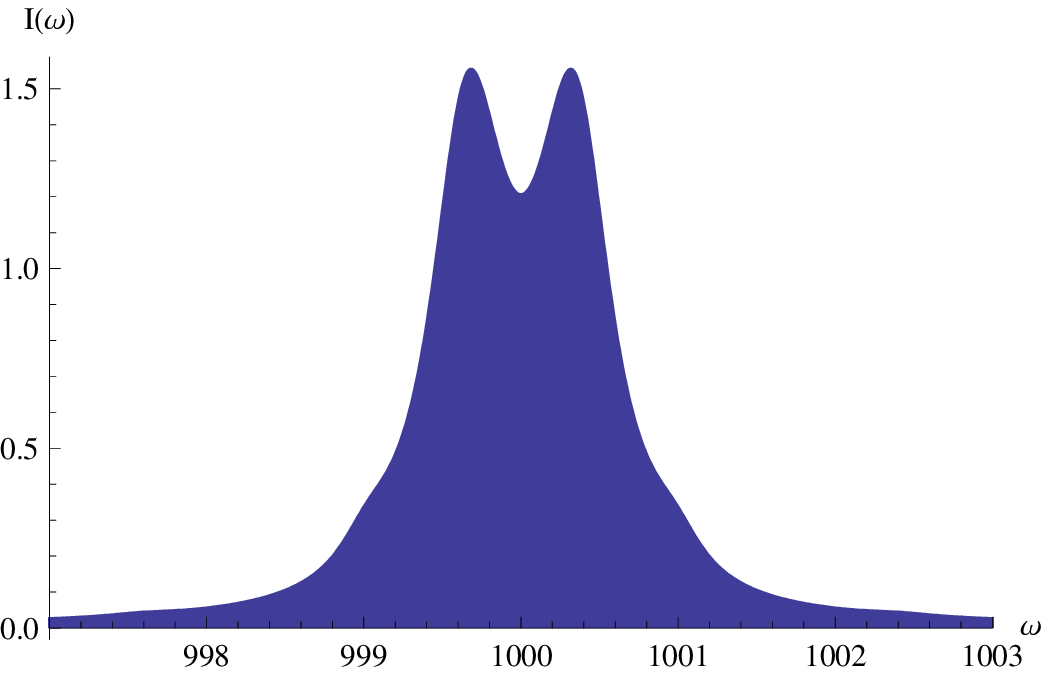}%
 \end{minipage}
\caption{\label{herbert} Emission spectrum of the system with different rates of
polaritonic pump $P_p$ for the parameters $\omega_X=1000$, 
$\kappa=0.1$, $\Delta=\gamma=P=0$. For this calculations we truncated equations
\ref{qrt} in $n_{\max}=20$}
\end{figure}

From the above discussion it is clear that the main effect of the term $P_p$ is
to cause an homogeneous broadening in the emission spectrum of the system. It
will also increase the intensity of the emitted light since it will
increase the population of highly excited states in the stationary limit,
equivalently the initial values of the two time operators. The two mentioned
effects can be seen in figure \ref{herbert} for resonance condition $\Delta=
0$ For low $P_p$ the peaks of the emission appear in $\omega_+=1001$ and
$\omega_-=999$ which are precisely the transition energies between the states
$\ket{1, \pm} $ and the vacuum $\ket{G0}$ ($\omega_X \pm g $). When the pumping
is increased new lines appear, this lines are associated with the energetic
transition between the states $\ket{2, \pm}$ and $\ket{1, \pm}$ which have
frequencies $\omega_X \pm g \pm \sqrt{2}g$ (for the parameters used they are
approximately $\omega_i \approx \{1002.4,1000.4,999.6,997.6\}$). When the
pumping is further increased the lines associated with the transition to
the vacuum become completely shadowed by the widened lines associated with the
transition $\ket{2, \pm} \longrightarrow \ket{1, \pm}$.

\subsection{Entanglement in the Stationary State}
\begin{figure}
   \begin{minipage}{0.3\textwidth}    
\centering 
$\sum_{i=3}^{n_{\max}} \rho_{Gn,Gn}+\rho_{Xn,Xn}$
\bigskip
  \end{minipage}
  \ \hfill
  \begin{minipage}{0.3\textwidth}    
\centering 
$\sum_{i=3}^{n_{\max}} \rho_{Gn,Gn}+\rho_{Xn,Xn}$
\bigskip
  \end{minipage}
  \ \hfill
  \begin{minipage}{0.3\textwidth}    
\centering 
$E(\rho)$
\bigskip
  \end{minipage}
\\
 \begin{minipage}{0.3\textwidth}
\includegraphics[width=1.0\textwidth]{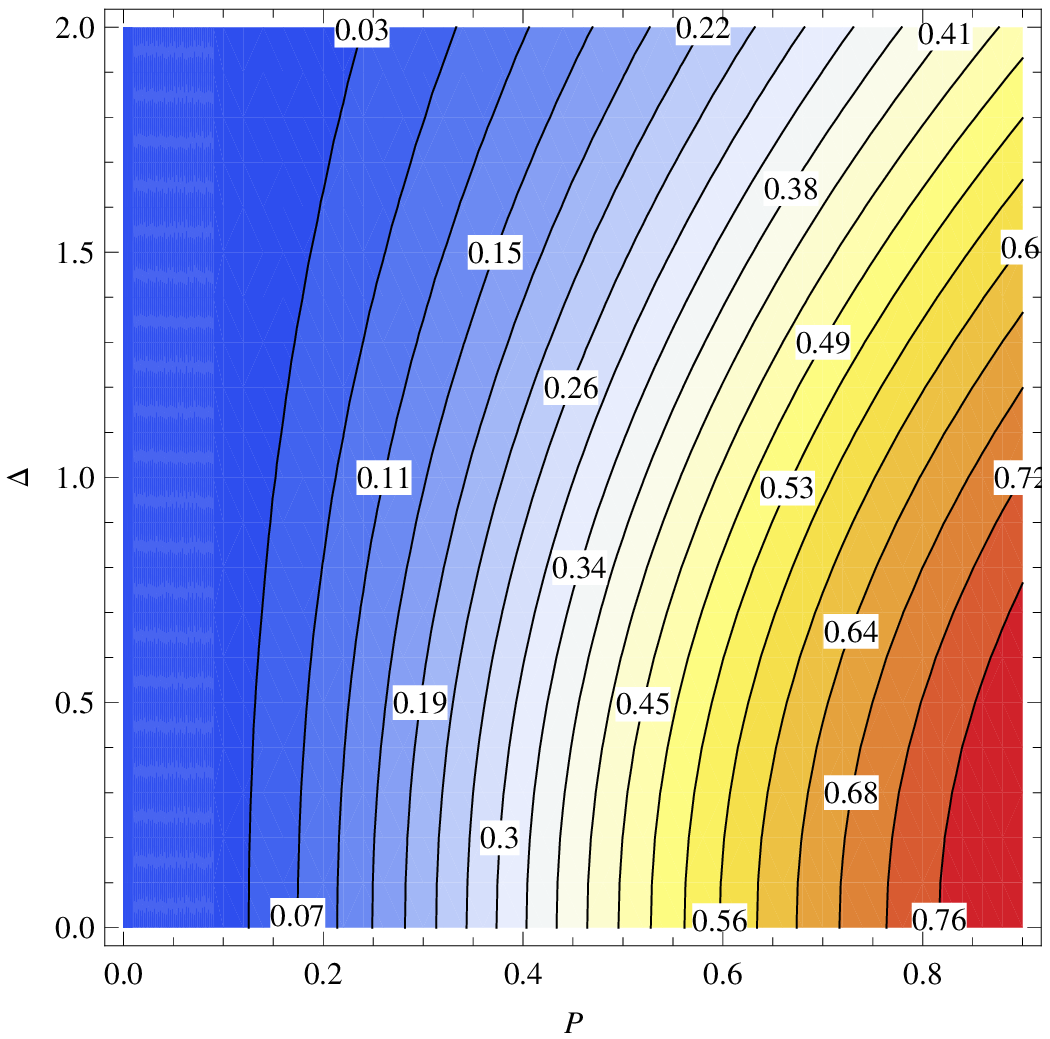}%
 \end{minipage}
\ \hfill
 \begin{minipage}{0.3\textwidth}
\includegraphics[width=1.0\textwidth]{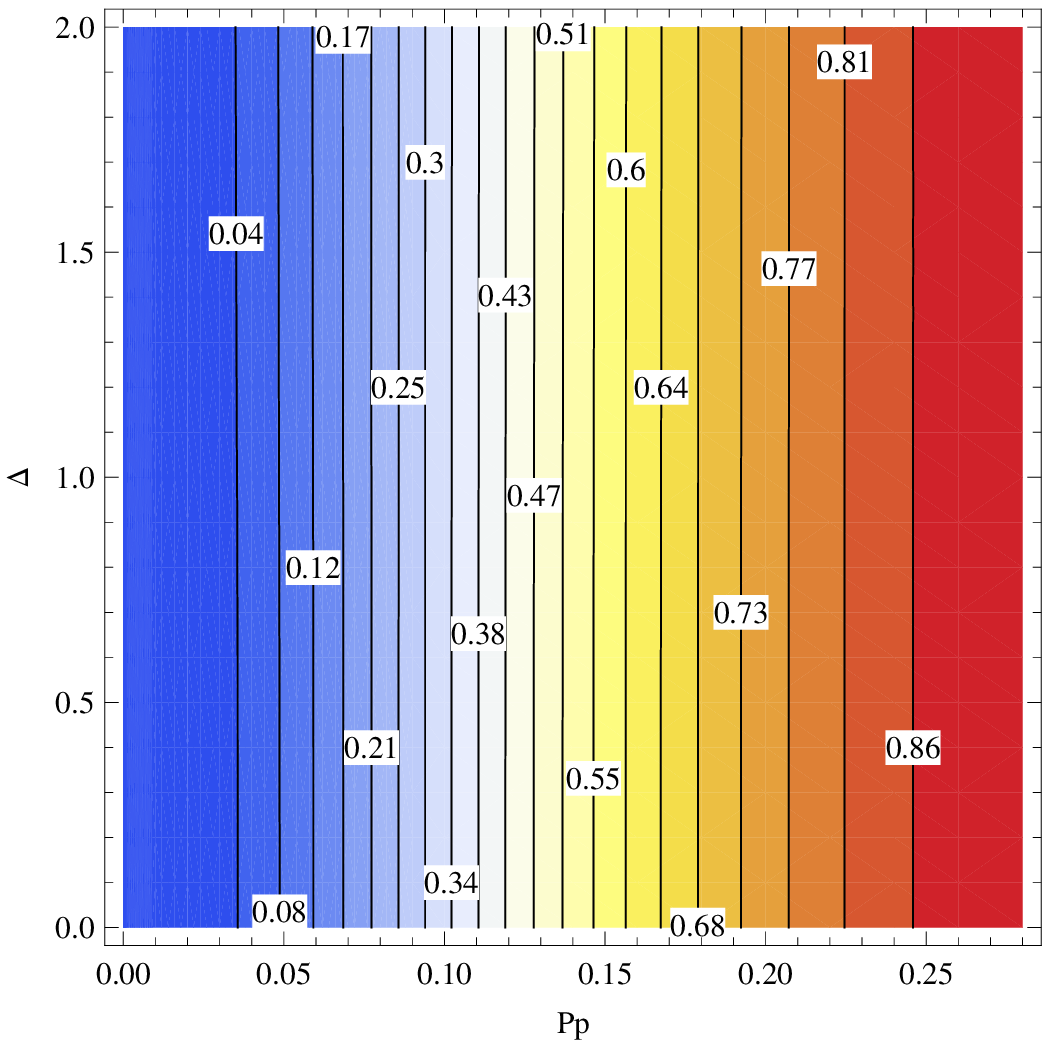}%
 \end{minipage}
\ \hfill
 \begin{minipage}{0.3\textwidth}
\includegraphics[width=1.0\textwidth]{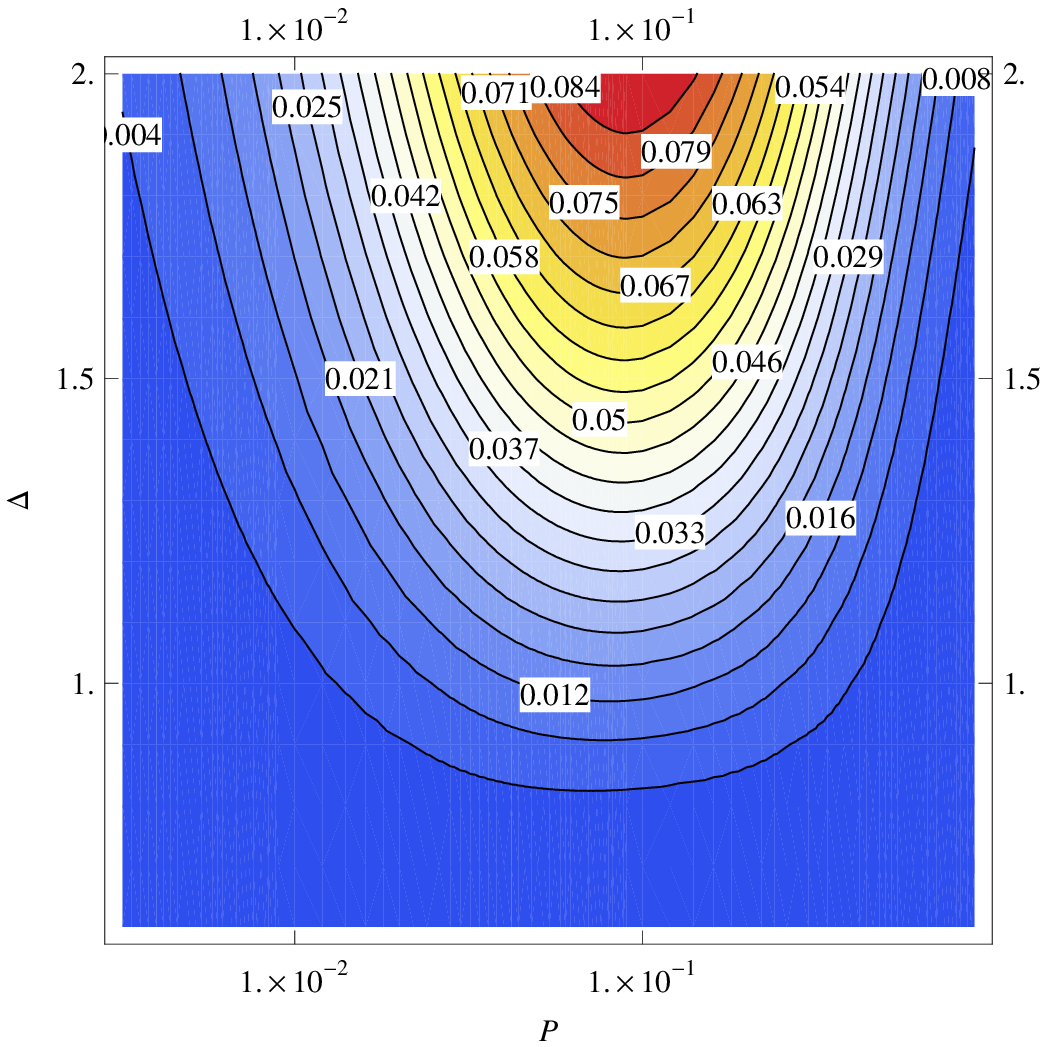}%
 \end{minipage}

\caption{\label{cutoff:ent} In the left and central panels we present the
fraction of the population of the density matrix that has more than two
photons as a function of the exciton or polariton pumping rates and the
detuning $\Delta$ for the parameters $\kappa=0.1$, $\gamma=0$, $g=1$. The
results do not change significatively if $\gamma=0.005$. The right panel
examines the entanglement measure $E(\rho)$ as a function of the exciton pumping
rate $P$ and the detuning $\Delta$ (The other parameters are the same of the
other two panels). It is seen that even in the stationary state one can have
entangled states, which is not case if one only pumps polaritons. The Fock
space was truncated in $n_{\max}=40$.}
\end{figure}

In this section, we use equation (\ref{ep}) to study the entanglement in the
stationary state. It is important to remember that the Peres criterion is a
necessary and sufficient condition for having entanglement when the dimensions
of the Hilbert spaces ($h_i$) of the subsystems considered are $\dim(h_1)=2$ and
$\dim(h_2)=2$ or $3$ \cite{Horodecki}, for larger dimensions it is only a
sufficient condition for having an entangled state. Under certain conditions the
Fock space of our system can be truncated in one or two photons. This
conditions are met when we consider
small enough pumping  rates ($P$ or $P_p$) as compared with the loss rates
($\kappa,\gamma$). This hypothesis is numerically confirmed by
figure \ref{cutoff:ent} where it is clearly seen that for $P$ or
$P_p< 0.1= \kappa+\gamma$ the population corresponding to states with more
than 2 photons
is very small or zero.
Then in the blue regions of figure \ref{cutoff:ent} we can obtain the stationary
state of the system considering only a Fock space of 2 photons. The exact
expressions for a maximum of 1 photon are presented in appendix \ref{appendix_B}. One 
can also find exact expressions for a maximum of 2 photons but they are
very cumbersome and are no presented here. In the cases where one considers
only the term $P_p$ ($P=\gamma=0$) one analytically finds that the state is
always separable (truncating in 1 or 2 photons), this was also numerically
confirmed using a larger basis ($n_{\max}=40$) for the parameter region
$10^{-4}<P_P<1$, $0<\Delta<2$, $\kappa=0.1$ and $\gamma=\{0,0.005\}$.
When  the effect of the excitonic pumping is considered we find
certain region of parameters where there are entangled states
(see the right panel of figure \ref{cutoff:ent}). This results highlight the
very special role of the exciton pumping term ($P$), because, its variation
might induce or reduce the degree of entanglement between the exciton and the field
mode.

\section{Conclusions}
In this work a mechanism for incoherent pumping of Polaritons
was proposed. The form of its matrix elements was derived and used in
section \ref{popelem} to obtain the dynamical equations necessary to propagate
the density operator of the system and obtain its emission spectrum.  The effects of
the new mechanism were compared with the effects of the exciton pumping. The physical origin of the new mechanism is still controversial and it is necessary further work in this direction. In the literature it has been considered an effective pump of excitons and photons. The fit that has been done in reference \cite{Finley} has shown that indeed both terms are able to account for a significant amount of the physics in such systems. Nevertheless using out off resonance excitation in quantum wells (QW) the group of J. Bloch obtains a polariton laser \cite{BlochJ}, that has been successfully modeled in reference \cite{Vera09b} including a effective polariton pumping term, that is essentially equivalent to our model. Additionally in reference \cite{Deng06} it has been shown that for QW an effective resonant pump to the lower polariton branch is a condition for thermalization of a BEC of polaritons.\\

In this work it was shown that the polariton pumping term is not able to cause a
positive population inversion and that above certain threshold drives the field
to a Poisson-like statistics where $g^2(\tau=0)=1$. It was shown that the term
$P_p$ is unable to change the dynamical regimes of the system, and that its
effect on the emission spectrum is twofold: it causes a homogeneous broadening of the peaks
and an increase in the integrated emission. Also in this work, it was examined
how the incoherent pumping mechanisms affects the entanglement of the exciton
and
the cavity in the stationary state. It was shown that the exciton pumping term
does not completely destroy entanglement as the polariton pumping does.
Finally, in section \ref{epp} a useful rule-of-thumb based on the Peres
criterion
was obtained to determine if the state of the exciton and the photons is
entangled.

\section*{\label{aknonw} Acknowledgments}
The authors acknowledge financial support from the Proyecto de
Sostenibilidad-GFAM from Universidad de Antioquia and Universidad Nacional de
Colombia. NQ is grateful to M.F. Suescum-Quezada for the help provided in the
diagramation of the figures.

\appendix

\section{Dynamics of the density matrix elements}
\label{appendix_A}
The dynamical equations for the populations and coherences in the bare basis are given by:
\begin{eqnarray}
\frac{d}{dt}\rho_{Gn,Gn}&=&-P \rho_{Gn,Gn}+\kappa  \left((n+1)
\rho_{Gn+1,Gn+1}-n \rho _{Gn,Gn}\right) \\
& &+P_p \left(\rho_{Gn-1,Gn-1}-2\rho _{Gn,Gn}+\rho _{Xn-2,Xn-2}\right) \nonumber
\\
& &+i g\sqrt{n}\left(\rho_{Gn,Xn-1}-\rho_{Xn-1,Gn}\right)+\gamma\rho_{Xn,Xn}
\nonumber \\
\frac{d}{dt}\rho_{Xn-1,Xn-1}&=&P \rho _{Gn-1,Gn-1} +\kappa  \left(n
\rho_{Xn,Xn}-(n-1) \rho_{Xn-1,Xn-1} \right)  \nonumber \\
& &+P_p \left(\rho_{Gn-1,Gn-1}+\rho_{Xn-2,Xn-2}-2 \rho _{Xn-1,Xn-1}\right)
\nonumber\\
& & +i g \sqrt{n}\left( \rho _{Xn-1,Gn}-\rho _{Gn,Xn-1}\right)-\gamma
\rho_{Xn-1,Xn-1} \nonumber \\
\frac{d}{dt}\rho_{Gn,Xn-1}&=&i g \sqrt{n} (\rho _{Gn,Gn}-\rho
_{Xn-1,Xn-1})+\kappa \sqrt{n(n+1)}  \rho_{Gn+1,Xn} \nonumber \\
& &+\left(-\frac{P}{2}-2P_p-\frac{\gamma }{2}+i \Delta -n \kappa
+\frac{\kappa}{2}\right) \rho _{Gn,Xn-1} \nonumber
\end{eqnarray}

\section{Stationary state truncating in one photon}
\label{appendix_B}

The populations and coherences of the stationary density operator, truncating the Fock space in one photon, are given by:
\begin{eqnarray}
\rho_{X0,G1}&=&\left(2 i g (P+4 P_p+\gamma -2 i \Delta +\kappa ) ((P+P_p)\kappa
-P_p \gamma )\right)/N\\
\rho_{G0,G0}&=&( 4 (P+2 P_p) (\gamma +\kappa ) (P+4 P_p+\gamma +\kappa )
   g^2\nonumber \\&+&P_p \gamma  (P+2 P_p+\gamma +\kappa ) \left(4 \Delta
   ^2+(P+4 P_p+\gamma +\kappa )^2\right))/C \nonumber \\
\rho_{X0,X0}&=&(4 (P+2 P_p) (\gamma +\kappa ) (P+4 P_p+\gamma +\kappa )
   g^2 \nonumber \\ &+&(P+P_p) \kappa  (P+2 P_p+\gamma +\kappa ) \left(4 \Delta
   ^2+(P+4 P_p+\gamma +\kappa )^2\right))/C \nonumber \\
\rho_{G1,G1}&=&(4 (P+2 P_p) (\gamma +\kappa ) (P+4 P_p+\gamma +\kappa )
   g^2 \nonumber \\&+&P_p \gamma  (P+2 P_p+\gamma +\kappa ) \left(4 \Delta
   ^2+(P+4 P_p+\gamma +\kappa )^2\right))/C \nonumber \\
\rho_{X1,X1}&=&(4 g^2 (P+4 P_p+\gamma +\kappa ) (P+2 P_p)^2 \nonumber \\& +&P_p
   (P+P_p) (P+2 P_p+\gamma +\kappa ) \left(4 \Delta ^2+(P+4
   P_p+\gamma +\kappa )^2\right))/C \nonumber\\
N&=&4 (P+2 P_p+\gamma +\kappa ) (P+4 P_p+\gamma +\kappa )
   g^2\nonumber\\&+&(P+P_p+\gamma ) (P_p+\kappa ) \left(4 \Delta ^2+(P+4
   P_p+\gamma +\kappa )^2\right)\nonumber\\
C&=&N \left(P+2 P_p+\gamma +\kappa\right) \nonumber
\end{eqnarray}

\section*{References}

\bibliographystyle{iopart-num}
\bibliography{pp}

\end{document}